\documentclass[aps,prb,twocolumn,longbibliography,superscriptaddress,floatfix]{revtex4-2}
\usepackage[utf8]{inputenc}
\usepackage{graphicx}
\usepackage{dcolumn}
\usepackage{bm}
\usepackage{amsmath}
\usepackage{amssymb}
\usepackage{amsthm}
\usepackage{mathtools}
\usepackage{mathrsfs}
\usepackage{float}
\usepackage{ulem}
\usepackage{epstopdf}
\usepackage{subfigure}
\usepackage{color}
\usepackage{newlfont}
\usepackage{epstopdf}
\usepackage{appendix}
\usepackage[breaklinks=true]{hyperref}
\usepackage{breakcites}
\usepackage{textcomp}
\usepackage{appendix}
\hypersetup{
     colorlinks   = true,
     linkcolor    = red,
     citecolor    = cyan,
     urlcolor     = magenta
     }

\usepackage{esvect}

\usepackage{soul}       

\graphicspath{{./Figures/}}

\begin{document}

\preprint{APS/123-QED}
\title{
Nonlinear skin breathing modes in one-dimensional nonreciprocal mechanical lattices
}

\author{Bertin Many Manda}
\email[]{bmany@tauex.tau.ac.il}
\affiliation{School of Mechanical Engineering, Tel Aviv University, Tel Aviv 69978, Israel}
\affiliation{%
  Laboratoire d’Acoustique de l’Universit\'e du Mans (LAUM), UMR 6613, Institut d'Acoustique - Graduate School (IA-GS), CNRS, Le Mans Universit\'e, Av. Olivier Messiaen, 72085 Le Mans, France 
}%



\date{\today}

\begin{abstract}

We investigate the interplay of nonreciprocity and nonlinearity in a one-dimensional nonlinear Klein-Gordon chain of classical oscillators coupled by asymmetric springs, akin to a mechanical analogue of the Hatano–Nelson model with onsite nonlinearity.
Using multiple-scale analysis, we show that families of nonlinear skin breathing modes—time-periodic, boundary-localized oscillations—emerge from their linear counterparts at any nonreciprocal strength. 
We derive an explicit nonlinear frequency shift for these families of nonlinear breathing modes, showing its dependence on amplitude, nonlinearity type, lattice size, and nonreciprocity, and we predict the emergence of genuine skin end breathers at the boundary once their oscillation frequency and higher harmonics enter the spectral gaps of the linear spectrum.
Numerical pseudo-arclength continuation confirms full families of solutions for both hardening and softening nonlinearities. 
Furthermore, the Floquet analysis shows that these modes can be either linearly stable or unstable, with Floquet eigenvectors exhibiting skin localization inherited from the asymmetric couplings. 
Our results extend the nonlinear non-Hermitian skin effect from stationary modes to intrinsically time-periodic excitations, providing a pathway to engineer and control breathing modes in nonreciprocal mechanical metamaterials.

\end{abstract}

\maketitle


\section{\label{sec:level1} Introduction}
Nonreciprocal mechanical metamaterials have recently attracted significant attention~\cite{BLLC2019,BMMFLTV2018,HHIAKMLST2020,WWM2022,ASPMCACBOPG2023,VGGSMC2024,JS2025,SDC2025,VGBVTCC2025}.
These metamaterials provide a versatile platform to explore non-Hermitian physics and wave control, as they can in general be effectively modeled by simple mass–spring systems in which the left- and right-going spring couplings differ from one another.
Among the most fundamental features of these systems is the non-Hermitian skin effect (NHSE), whereby all the normal modes collapse at one edge of the system under open boundary conditions (OBCs)~\cite{KEBB2018,LT2019,YW2018,OKSS2020,OS2023,WC2023,ZZCLC2023}.
The NHSE was first demonstrated in experiment in mechanical systems~\cite{BLLC2019,GBVC2020} before finding analogues in electronics~\cite{LSMZYWJJZ2021,JMAFS2025}, acoustics~\cite{ZYGGCYCXLJYSCZ2021,MAPPA2023}, optics~\cite{WKHHSGTS2020,PVRRVF2025}, and other phononic settings~\cite{LDHKL2025}.
These developments provide strong motivation for a deeper and systematic investigation of nonreciprocal mechanical systems.

Nevertheless, research on the interplay between the NHSE and nonlinearity has largely overlooked mechanical systems, focusing instead on photonic lattices like the Hatano–Nelson (HN) model with onsite nonlinearity~\cite{Y2021,E2022c,KMM2023,JCZL2023,G2024,MCKA2024,WWLQZLLL2025}.
Through these studies it was found that nonlinear NHSE persists for finite-amplitude waves.
More specifically, perturbation methods clarified in theory and numerical simulations the emergence of nonlinear skin modes from their linear counterparts as the nonlinearity grows~\cite{KMM2023,JCZL2023,MCKA2024}.
However, the particularity of these modes is that they are stationary, a property inherent from systems governed by first-order differential equations such as $i\dot{y}=Hy$.
Indeed, solutions in such systems are of the form $y(t)=U e^{-i\Omega t}$ reducing their dynamics to the eigenvalue problem $Hu=\Omega u$. 
It follows that, the field profile $\lvert y(t)\rvert = U$ becomes time-independent~\cite{KMM2023,JCZL2023,MCKA2024}.

\begin{figure}[!t]
    \centering
    \includegraphics[width=\columnwidth]{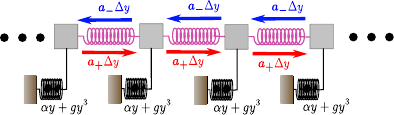}
    \caption{Schematic of a chain of classical oscillators interconnected with asymmetric elastic spring couplings. The relation between the force, $F$, and the spring's elongation, $\Delta y$ is given by $F_{\text{Right}\rightarrow \text{Left}} = a_{+}\Delta y$ (red) and $F_{\text{Left}\rightarrow \text{Right}} = a_{-}\Delta y$ (blue), where $a_{\pm} = 1 \mp \gamma$ are the elastic constants and $\gamma$ measures the nonreciprocal strength.
    In addition, the onsite restoring force is given by $F = \alpha y + g y^3$ (black), where $\alpha$ and $g$ are the onsite elastic and nonlinear spring coefficients, respectively, and $y$ denotes the oscillator displacement.
    }
    \label{fig:nonreciprocal_KG_chain}
\end{figure}

In mechanical lattices, however, the dynamics is governed by second-order equations like, $\ddot{y} = H y$.
It follows that when looking for solutions of the form, $y(t) = U e^{-i\Omega t}$ yields $\ddot{y} = -\Omega^{2} y$ with $HU = -\Omega^2 U$.
Thus, the system dynamics necessarily contains $e^{\pm i \Omega t}$ terms leading to a general form of the solution with two components $y(t) = U^\prime \cos(\Omega t) + U^{\prime\prime}\sin(\Omega t)$, whose amplitudes $\lvert y(t)\rvert$ oscillate in time.
It follows that the nonlinear modes in mechanical systems are inherently time-periodic (like discrete breathers~\cite{A1997,FW1998,FG2008,CK2018}) rather than stationary.
This difference raises several fundamental questions.
Do nonlinear skin breathing (oscillating) modes exist in nonreciprocal mechanical lattices? 
If so, how do their frequencies and envelopes depend on control parameters like the nonlinearity, nonreciprocity, and lattice size?

We address these questions by studying a one-dimensional (1D) chain of nonlinear classical oscillators connected through asymmetric spring couplings, referred to as the nonlinear nonreciprocal Klein-Gordon (KG) chain.
This model originates from the mechanical analogue of the celebrated Hatano–Nelson (HN) chain~\cite{HN1996,HN1998} with a cubic onsite nonlinearity~\cite{MCKA2024}. 
Mechanical implementations of closely related systems already exist, including nonreciprocal variants of the discrete sine–Gordon model~\cite{VGGSMC2024,VGBVTCC2025} (see also~\cite{JMAFS2025}).
Notably, Ref.~\cite{VGBVTCC2025} investigates, in the presence of dissipation, traveling breathers within the bulk of the lattice.
By contrast, the present work focuses on nonlinear normal modes that originate from the linear skin modes. 
It follows that by definition, these states are standing, time-periodic, and localized at the edge selected by the NHSE.

In this paper, we demonstrate that the nonlinear NHSE extends to time-periodic waves in nonreciprocal lattices. 
Using multiple-scale analysis and numerical continuation methods, we show that nonlinear skin breathing modes emerge from their linear counterparts in arbitrarily large but finite lattices, with amplitudes that oscillate in time for both softening and stiffening nonlinearities.
We further investigate the linear stability of these modes through Floquet analysis, finding that they can be linearly stable or unstable depending on the system's control parameters. 
Interestingly, the associated Floquet eigenvectors themselves exhibit skin profiles inherited from the spatial asymmetry of the nonreciprocal couplings.

The remainder of the paper is organized as follows. In Sec.~\ref{sec:skin_breathers} we present our nonlinear nonreciprocal KG chain and study its spectral properties in both the linear and nonlinear regimes. 
Section~\ref{sec:stability_breathers} analyzes the linear stability of the obtained nonlinear modes via Floquet analysis and shows representative examples of their dynamics. 
Finally, in Sec.~\ref{sec:conclusion_breathers}, we summarize our work and outline several directions for future research.

\section{\label{sec:skin_breathers}Nonlinear skin breathing modes}

\begin{figure}[!tb]
    \centering
    \includegraphics[width=\columnwidth]{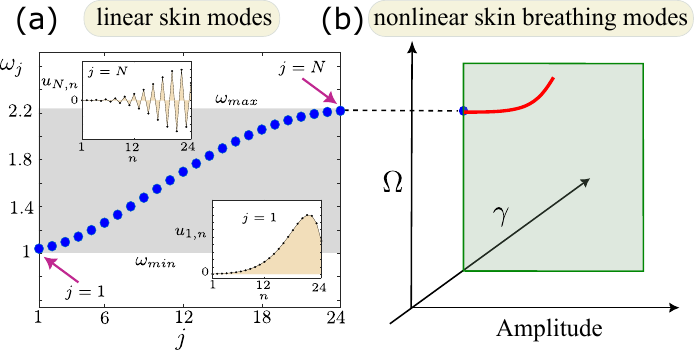}
    \caption{Dependence of the numerically obtained linear frequency, $\omega_j$ against the wave number $j$ for the nonreciprocal KG chain of Fig.~\ref{fig:nonreciprocal_KG_chain} with $N=24$, $\gamma =0.25$, $\alpha =1$, $g=0$, and fixed boundary conditions.
    The lower and upper frequency cutoffs are $\omega_{min} = (2 + \alpha - 2\sqrt{1-\gamma^2})^{1/2}$ and $\omega_{max} = (2 + \alpha + 2\sqrt{1-\gamma^2})^{1/2}$ respectively (see text for details).
    The bottom and top insets show the right eigenvectors, $u_{j,n}$ with wave numbers $j=1$ and $j=N$.
    (b) Procedure for the computation of a family of nonlinear skin breathing modes emerging from their linear counterparts.
    }
    \label{fig:skin_modes_and_procedure}
\end{figure}

We consider the nonlinear nonreciprocal KG chain.
As shown in Fig.~\ref{fig:nonreciprocal_KG_chain}, this lattice is composed of a one-dimensional arrangement of nonlinear classical oscillators, coupled to their nearest neighbors through identical elastic springs with asymmetric couplings.
It follows that the motion of the oscillator at site with label $n$ is generated through (see also~\cite{JMAFS2025,VGBVTCC2025})
\begin{equation}
    \ddot{y}_n = a_{-} y_{n-1} - 2 y_n + a_{+} y_{n+1} - \alpha y_n - g y_n^3,
    \label{eq:eq_motion}
\end{equation}
where $y_n$ and $\dot{y}_n = dy_n/dt$  are its generalized displacement and conjugate momentum with $\vec{y} = \left(y_1, y_2, \ldots, y_N, \dot{y}_1, \dot{y}_2, \ldots, \dot{y}_N\right)^T$ [$\{^T\}$ stands for the vector transpose].
The $\alpha$ and $g$ are the onsite elastic and nonlinear coefficients, with positive values of $g$ giving rise to hardening-type nonlinearities, whereas negative ones lead to softening-type nonlinearities.
We set the asymmetric elastic coefficients $a_{\pm} = 1 \mp \gamma$, following modeling in experiments in Ref.~\cite{JMAFS2025}, with $\gamma$ defining the nonreciprocal strength.
Furthermore, we consider a chain of $N$ sites with fixed boundary conditions imposed at both ends, namely $y_0 = y_{N+1} = 0$.

 \begin{figure*}[!t]
    \centering
    \includegraphics[width=\textwidth]{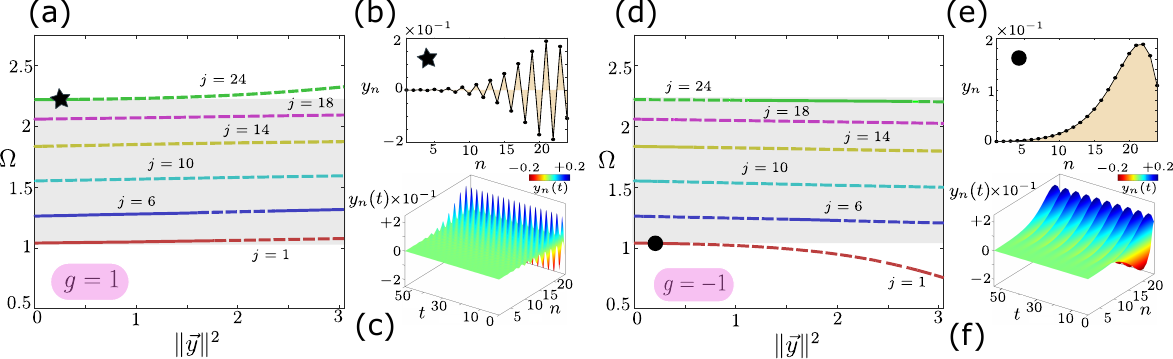}
    \caption{Dependence of the nonlinear frequency $\Omega$ on the amplitude $\lVert \vec{y} \rVert^{2}$ for families of nonlinear skin breathing modes emerging from the linear skin modes with $j = 1$ (red), $j = 6$ (blue),  $j = 10$ (cyan), $j = 14$ (yellow), $j=18$ (purple) and $j = 24$ (green), shown from the bottom to the top curves.
    The dashed portions of the curves indicate unstable modes, while the solid portions correspond to stable ones (this representation is qualitative).
    The parameters used in the computations are $N = 24$, $\alpha = 1$, $\gamma = 0.25$, and $g = +1$ (hardening). The numerical continuation is stopped when $\lVert \vec{y} \rVert^{2} = 3$, which roughly corresponds to weak and moderate nonlinear strengths.
    (b) Profile of a representative nonlinear skin breathing mode with $\lVert \vec{y}\rVert^{2} = 0.2$, belonging to the family with $j=24$, see star symbol.
    Note that its linear counterpart is shown in the upper inset of Fig.~\ref{fig:skin_modes_and_procedure}(a).
    (c) Time evolution of the state in (b) used as the initial condition, clearly demonstrating an amplitude that oscillates in time.
    (d) Same as in (a), but for $g = -1$ (softening).
    (e) Same as in (b), but for the softening case shown in (d), see dot symbol.
    Note that its linear counterpart is depicted in the lower inset of Fig.~\ref{fig:skin_modes_and_procedure}(a).
    (f) Same as in (c), but using the state in (e) as the initial condition.
    }
    \label{fig:nonlinear_continuation_results}
\end{figure*}

Notably, in the linear regime of small amplitude waves, we look for harmonic solutions of the form  $y_n = u_n e^{i\omega t}$ where $u_n$ is an amplitude, $\omega$ stands for a real frequency and $i^2=-1$.
Consequently the equations of motion reduces to an eigenvalue problem
\begin{center}
    \begin{equation}
        \omega^2 \vec{u} = D \vec{u},
        \label{eq:dynamical_matrix}
    \end{equation}
\end{center}
where $\vec{u}=(u_1, u_2, \ldots, u_{N-1}, u_N)^T$, assuming all the momenta are zero and $D$ is a tri-diagonal dynamical matrix with real entries as shown in the Appendix~\ref{app_sec:matrices}.
In particular, the choice of $a_{\pm} = 1 \mp \gamma$ ensures that the diagonal elements of $D$ remain fixed,  $D_{n,n} = 2 + \alpha$ ($n=1, 2, \ldots, N$).
Consequently, the symmetries of $\omega^2$ are conserved regardless of the nonreciprocal strength $\gamma$.

The resulting squared dispersion relation
\begin{equation}
    \omega_j^2 = 2 + \alpha - 2\sqrt{1-\gamma^2}\cos \left( \frac{j\pi}{N+1} \right),
    \label{eq:linear_square_frequency}
\end{equation}
with $j = 1, 2, \ldots, N$, is symmetric with respect to the mid-squared frequency $\omega_0^2 = 2+\alpha$.
In addition, it exhibits a lower gap below $\omega_{min}^2 = 2 + \alpha - 2\sqrt{1-\gamma^2}$ and an upper one above $\omega_{max}^2=2 + \alpha + 2\sqrt{1-\gamma^2}$.
Further, the $\omega_{min}^2$ (resp.  $\omega_{max}^2$) monotonically increases (resp. decreases) with growing nonreciprocity, $\gamma$.
Therefore, in the limit of a fully nonreciprocal chain ($\gamma = 1$), all the squared frequencies collapse to $\omega_j^2 = \omega_0^2$.

We take a chain of $N=24$ sites, $\alpha = 1$ with $\gamma =0.25$ and perform the numerical diagonalization of the dynamical matrix, $D$.
Figure~\ref{fig:skin_modes_and_procedure}(a) shows the obtained frequencies $\omega_j$ against the wave number $j$.
Clearly even in case of a small lattice size, we confirm the presence of a lower band gap, since $\omega_{min}=  1.04$ is much larger than the average frequency spacing, $\langle \omega_{j+1} - \omega_j\rangle \approx 0.05$.

Furthermore, as the dynamical matrix is non-Hermitian (Appendix~\ref{app_subsec:dynamical_matrix}), we can also define a left eigenvalue problem $\omega^2\vec{v}=D^T\vec{v}$, with $\vec{v}=(v_1, v_2, \ldots, v_{N-1}, v_N)^T$.
The right and left eigenvectors associated with a frequency $\omega_j$, 
\begin{equation}
    u_{j,n} =  r^{\frac{n}{2}}\sin \left(\frac{nj\pi}{N+1}\right),~~ v_{j,n} =  r^{-\frac{n}{2}}\sin \left(\frac{nj\pi}{N+1}\right)
    \label{eq:normal_modes}
\end{equation}
are spatially asymmetric, with $r=a_{-}/a_{+} = (1+\gamma)/(1-\gamma)$.
They are exponentially localized and respectively located at the right and left ends of the chain whenever $\gamma \in \left(0, 1\right]$.
As such, they are referred to as {\it skin modes}, and are exactly what induces the NHSE.
The lower and upper insets of Fig.~\ref{fig:skin_modes_and_procedure}(a) show representative right eigenvectors, $u_{j,n}$, corresponding to $j=1$ and $j=N$, obtained via the numerical diagonalization discussed above.
It is these right eigenvectors we will generalize in the nonlinear regime.

We now address whether families of nonlinear normal modes satisfying Eq.~\eqref{eq:eq_motion} can emerge from the linear skin modes as their amplitudes increase.
A schematic illustration of this procedure and the expected outcome is shown in Fig.~\ref{fig:skin_modes_and_procedure}(b), where the parameter $\gamma$ is fixed while the frequency, $\Omega$ is treated as a free parameter.
We apply the method of multiple scales, seeking solutions of the form~\cite[Chap.~6]{NM1979}
\begin{equation}
    y_n = \epsilon \left[ w_nA(T) e^{i\Omega t} + w_n^\star A^\star (T)  e^{-i\Omega t} + \epsilon^2 y_n^{(3)} + \ldots \right],
\label{eq:ansatz_hermitian}
\end{equation}
where $\epsilon \ll 1$ and the $\{ ^\star\}$ stands for the complex conjugate operation.
In addition, the $t$ and $T=\epsilon^2t$ are fast and slow time variables.
We find that at order $\epsilon$, $\Omega^2 w_n = \sum_m D_{n, m}w_m$.
Consequently, $\Omega = \omega_j$ and $w_n = u_{j,n}$, indicating that the ansatz correctly reproduces the linear frequencies and the associated skin modes in the small amplitude limit.
Proceeding to order $\epsilon^3$, the condition for no resonant terms leads to the amplitude equation
\begin{equation}
    2 i \omega_j \frac{\partial A}{\partial T} + 3 \mu \lvert A\rvert^2 A = 0,
    \label{eq:amplitude_equation}
\end{equation} 
with $\mu = 2g\left(N+1\right)^{-1}\left(\sum_{n=1}^{N} v_{j,n} u_{j,n}^3\right)$.
It follows that,
\begin{equation}
   A(\epsilon^2 t) = A_0 \exp \left(i \epsilon^2 t \frac{3\mu \lvert A_0 \rvert^2}{2\omega}\right).
   \label{eq:breathers_expression_in_hermitian}
\end{equation}
The details of these derivations are reported in Appendix~\ref{app_sec:muliple_scale}.

Let us now express $\epsilon$ in terms of a macroscopic parameter controlling the nonlinearity in our system, namely the squared maximum displacement, $\lVert \vec{y}\rVert^{2}$ (hereafter referred to as the total amplitude), which is a constant coefficient.
Indeed, we see that the complex amplitude equation [Eq.~\eqref{eq:amplitude_equation}] is a Hermitian nonlinear Schr\"odinger equation, which possesses conserved quantities; in particular, $\lvert A(T)\rvert^2 = \lvert A_0\rvert^2$, see Appendix~\ref{app_sec:conservation}.
It follows that, at the phase of the solution for which the momenta vanish ($\dot{y}_n = 0$), Eq.~\eqref{eq:ansatz_hermitian} yields $4 \epsilon^2 \lvert A_0 \rvert^2 = \lVert \vec{y}\rVert^2 / \lVert \vec{u}\rVert^2$ when retaining only the first order in $\epsilon$.
Consequently,
\begin{equation}
    y_n (t) = \sqrt{S} \frac{u_n}{\lVert \vec{u}\rVert} \cos \left( \Omega t\right), \quad S = \lVert \vec{y}\rVert^2,
    \label{eq:breather_shape}
\end{equation}
with
\begin{equation}
    \Omega_j = \omega_j + \frac{3 S g }{4(N+1)\omega_j} \left[\dfrac{\displaystyle \sum_{n=1}^{N} r^n \sin^4 \left(\frac{nj\pi}{N+1} \right)}{\displaystyle \sum_{n=1}^N r^n \sin^2 \left(\frac{nj\pi}{N+1} \right)} \right].
    \label{eq:breather_frequency}
\end{equation}
It is worth emphasizing that from Eq.~\eqref{eq:breathers_expression_in_hermitian} to Eq.~\eqref{eq:breather_frequency}, we replaced $u_{j,n}$ and $v_{j,n}$ by their respective expressions. 
Further, the summations appearing in Eq.~\eqref{eq:breather_frequency} can be evaluated in closed forms leading to algebraically cumbersome expressions which we do not substitute. 

Equations~\eqref{eq:breather_shape} and~\eqref{eq:breather_frequency} demonstrate that nonlinear skin breathing modes stem from the linear skin modes given a nonlinear correction to their frequencies, $\omega_j$, quantified by $\Omega_j - \omega_j$.
Interestingly, this shift depends on the total amplitude $\lVert \vec{y}\rVert^2$, the nonreciprocal strength, $\gamma$, and the lattice size $N$.
Overall, for softening nonlinearities ($g<0$), $\Omega$ typically decreases approximately linearly with $\lVert \vec{y}\rVert^2$, whereas for hardening nonlinearities ($g>0$) the trend is reversed.

Lets us now perform numerical verifications of the theory above.
We implement a pseudo-arclength solver~\cite{DKK1991,KPGV2009,PVSKG2009} to numerically find families of exact nonlinear solutions of Eq.~\eqref{eq:eq_motion}. 
The initial point for the pseudo–arclength solver is obtained using a shooting method, which searches for a small–amplitude exact solution with frequency $\Omega$, using the corresponding linear skin mode as an initial guess~\cite{KPGV2009,PVSKG2009}.
Note that the resulting nonlinear modes are obtained with phases chosen such that their momenta satisfy $\dot{y}_n = 0$.
For the direct numerical simulations of the equations of motion, we use a Runge–Kutta scheme of order $8$ based on the Dormand–Prince method (DOP$853$), with a time step $\Delta t = 0.05$ and a one-step tolerance of $10^{-11}$~\cite{HNW1993,DMMS2019,freelyDOP853}. 
Hereafter, we consider $N = 24$, $\alpha = 1$, $\gamma = 0.25$ and terminate the pseudo-arclength continuation at $\lVert \vec{y} \rVert^{2} = 3$ so as to avoid intricate bifurcation structures that emerge at large nonlinearities, which are beyond the scope of this work.

\begin{figure}[!tb]
    \centering
    \includegraphics[width=\columnwidth]{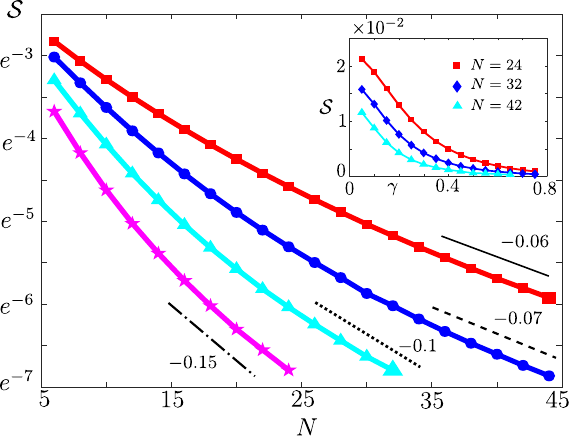}
    \caption{
        Dependence of the nonlinear frequency sensitivity factor $\mathcal{S} = \partial \widetilde{\Omega} / \partial S$ against the lattice size $N$ with $\widetilde{\Omega} = (\Omega - \omega)/\omega$ and $S = \lVert \vec{y} \rVert^{2}$.
        We compute $\mathcal{S}$ for fixed values $\gamma = 0.25$ (red squares), $\gamma = 0.4$ (blue dots), $\gamma = 0.55$ (cyan triangles), and $\gamma = 0.7$ (purple stars). 
        The derivatives are evaluated using a two-point finite difference at $S = 0.01$ and $0$~\cite{NOTE0001}.
        The straight solid, dashed, dotted, and dashed–dotted lines guide the eye to $\mathcal{S}\sim e^{-qN}$ with $q=0.06$, $0.07$, $0.10$, and $0.15$ (see text for details).
        Inset: Dependence of $\mathcal{S}$ on $\gamma$ for three lattice sizes, $N = 24$, $N = 32$, and $N = 42$, shown as the blue, red, and cyan dotted symbols. 
        We find that $\mathcal{S}(\gamma)\sim r^{-2}$, $r^{-2.5}$ and $r^{-3}$ for $N = 24$, $N = 32$, and $N = 42$ respectively.
        In all cases, the analytical predictions show as solid curves overlapping with the numerical results, see also Fig.~\ref{fig_app:freq_amplitude_theo_numerics}.
    }
    \label{fig:wb_against_size}
\end{figure}

Figure~\ref{fig:nonlinear_continuation_results} shows the dependence of the nonlinear frequency $\Omega$ against the amplitude $\lVert \vec{y} \rVert^2$, focusing on representative modes with $j=1$ (red), $j=6$ (blue), $j=10$ (cyan), $j=14$ (yellow), $j=18$ (purple) and $j=24$ (green) within the weak and moderate nonlinear regimes.
We first consider the case of hardening nonlinearity with $g = 1$.
Figure~\ref{fig:nonlinear_continuation_results}(a) depicts families of nonlinear breathing modes emerging from their linear counterparts, with $\Omega$ monotonically increasing as the amplitude grows. 
In addition, within $0 <\lVert \vec{y}\rVert < 3$, the $\Omega$ of the families with $j = 1,~6,~10,~14$, and $18$ remain well inside the linear spectrum [gray region].
In contrast, for the family with $j = 24$, both the nonlinear frequency $\Omega$ and its higher harmonics lie outside the linear spectrum, giving rise to genuine skin end breathers.
We compare in Fig.~\ref{fig_app:freq_amplitude_theo_numerics} of Appendix~\ref{app_sec:muliple_scale}, the analytical results (black curves) obtained using Eq.~\eqref{eq:breather_frequency} with the ones from the numerical continuation and see excellent agreements.

Moreover, for illustrative purposes, Fig.~\ref{fig:nonlinear_continuation_results}(b) shows the profile of a representative solution of the family arising from the mode with $j = 24$ and $\lVert \vec{y} \rVert^2 = 0.2$ [star symbol in Fig.~\ref{fig:nonlinear_continuation_results}(a)].
The displacements exhibit a skinny profile, similar to that of the linear skin modes shown in the upper inset of Fig.~\ref{fig:skin_modes_and_procedure}(a).
It is also important to highlight the spatio-temporal dynamics of these nonlinear skin breathing modes, which distinguish themselves from stationary solutions studied in nonlinear HN lattices~\cite{Y2021,KMM2023,MCKA2024}.
In particular, using the representative example of Fig.~\ref{fig:nonlinear_continuation_results}(b) [i.e., $j = 24$, $\lVert \vec{y} \rVert = 0.2$, $g = 1$] as the initial condition for our numerical integration, we observe that the displacements $y_n(t)$ exhibit amplitudes that oscillates in time with period $T_b = 2\pi/\Omega =  2.83$ up to the final time, $t \approx 50$, Fig.~\ref{fig:nonlinear_continuation_results}(c).

Similarly, Fig.~\ref{fig:nonlinear_continuation_results}(d) presents the numerical continuation results in case of a softening nonlinearity with $g = -1$, using the same parameters as above.
We see that the nonlinear frequency $\Omega$ decreases as $\lVert \vec{y} \rVert^{2}$ grows. 
Besides, in cases of families with $j = 2,~6,~10,~14,~18$, and $24$, the nonlinear frequencies remain well within the linear spectrum (gray region) when $\lVert \vec{y}\rVert^2 \rightarrow 3$.
In contrast, the family emerging from the mode with $j = 1$ has its frequency and higher harmonics lying inside the lower band gap, thereby leading to skin end breathers also in the softening case.
We show in Fig.~\ref{fig:nonlinear_continuation_results}(e) a representative profile of these families of nonlinear breathing modes, focusing on the case with $j=1$ and $\lVert \vec{y} \rVert^2 = 0.2$ [see dot symbol in Fig.~\ref{fig:nonlinear_continuation_results}(d)].
Its shape also depicts a skinny profile, similar to skin mode of its linearized limit [see lower inset of Fig.~\ref{fig:skin_modes_and_procedure}(a)].
Furthermore, the time evolution of this state clearly reveals a nonlinear mode with amplitudes  at every lattice's site oscillating in time with period $T_b = 2\pi/\Omega = 6.06$.

\begin{figure}[!tb]
    \centering
    \includegraphics[width=\columnwidth]{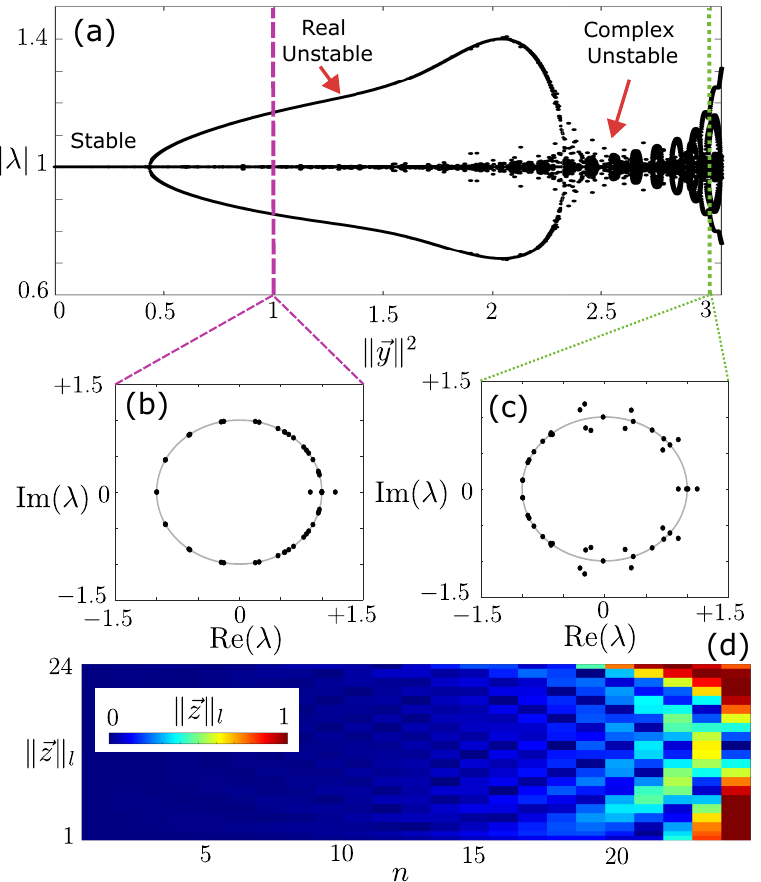}
    \caption{
    (a) Dependence of $\lvert \lambda \rvert$ against the amplitude $\lVert \vec{y}\rVert^{2}$ for the representative family emerging from the skin mode of $j=1$ with $N = 24$, $\alpha = 1$, $\gamma = 0.25$, and $g = -1$ [see also the red curve in Fig.~\ref{fig:nonlinear_continuation_results}(d)].
    Values of $\max \lvert \lambda \rvert \neq 1$ signal instability.~
    (a)–(b) Floquet eigenvalues $\lambda$ in the complex plane for the representative nonlinear breathing modes with amplitudes $\lVert \vec{y}\rVert = 1$ [purple dashed horizontal line in (c)] and $\lVert \vec{y}\rVert^2 = 3$ [green dotted horizontal line in (c)].
    Eigenvalues lying outside the unit circle (gray circle) along the real axis indicate real instabilities, while those outside the circle off the real axis correspond to complex instabilities.
    (d) Floquet eigenvectors $\lVert \vec{z}_l\rVert $ corresponding to the eigenvalues shown in (b).
    The vectors $\vec{z}_l$ are sorted in increasing order of $\lvert \lambda \rvert$.
    For clarity, each eigenvector $\lVert \vec{z}_l \rVert$ is normalized by $\max \lvert z_{l,n} \rvert$.
    }
    \label{fig:floquet_analysis}
\end{figure}

Let us now examine the dependence of the frequency of these families of nonlinear breathing modes on the lattice size $N$ and the nonreciprocity parameter $\gamma$. 
In this context, it is convenient to introduce
\begin{equation}
\mathcal{S} =  \left\lvert \frac{\partial \widetilde{\Omega}}{\partial S}\right\rvert  = \frac{3 \lvert g \rvert }{4(N+1)\omega_j^2} \left[\dfrac{\displaystyle \sum_{n=1}^{N} r^n \sin^4 \left(\frac{nj\pi}{N+1} \right)}{\displaystyle \sum_{n=1}^N r^n \sin^2 \left(\frac{nj\pi}{N+1} \right)} \right],
\label{eq:sensitivity_factor}
\end{equation}
where $\widetilde{\Omega} = (\Omega - \omega_j)/\omega_j$ is the rescaled nonlinear frequency.
This quantity measures how strongly the frequency changes with increasing amplitude, and can therefore be regarded as a frequency sensitivity factor, see e.g.~\cite{BM2024}.
Although an explicit analytical extraction of the dependencies of this sensitivity factor is challenging, the geometric-series nature of the summations involved in $\mathcal{S}$ strongly suggests power-law scaling with respect to both $N$ and $\gamma$.

We numerically compute the dependence of this frequency sensitivity factor, $\mathcal{S}$ [Eq.~\eqref{eq:sensitivity_factor}] against the lattice size, $N$ in case of the hardening nonlinearitly with $g = 1$.
Note that the first derivative is evaluated using a two points finite difference formula in the interval $\lVert \vec{y}\rVert^2 \in (0, 0.01]$~\cite{NOTE0001}.
Figure~\ref{fig:wb_against_size} shows the results of these computations for the family of nonlinear skin breathing modes with $j=1$ considering four different values of the nonreciprocity, $\gamma = 0.25$, $\gamma = 0.4$, $\gamma = 0.55$, and $\gamma = 0.7$, corresponding respectively to the red squares, blue dots, cyan triangles, and purple stars in Fig.~\ref{fig:wb_against_size}.
We find that for all values of $\gamma$, the $\mathcal{S}$ tends to asymptotically decay as the lattice size growth.
Assuming the scaling $\mathcal{S}\sim e^{-qN}$ reveals that the exponent $q$ depends on $\gamma$, i.e., $q=q(\gamma)$.
Interestingly, we find that $q$ increases with increasing $\gamma$.
Specifically, for $\gamma=0.25,~0.4,~0.55$ and $0.7$, we obtain $q(\gamma)=0.06,~0.07,~0.1$ and $0.15$, respectively, as indicated by the black solid, dashed, dotted, and dashed–dotted lines in Fig.~\ref{fig:wb_against_size}.
The exponent $q$ is extracted by fitting the last five numerical data points.

Moreover, the inset of Fig.~\ref{fig:wb_against_size} depicts similar computations as above, but varying this time the nonreciprocal strength $\gamma$ at fixed lattice sizes, $N=24$, $N=32$ and $N=42$ respectively the red squares, blue diamonds and cyan triangles.
Following similar arguments as above, we expect $\mathcal{S}\sim r^{p}$, with $p= p(N)$.
By numerical fitting the sensitivity using the power law $\mathcal{S} = \sigma r^{-p}$, we find that $\mathcal{S}(\gamma) \sim r^{-2}$, $ r^{-2.5}$, and $r^{-3}$ for $N=24$, $N=32$, and $N=42$, respectively~\cite{QQQQQQQ}.
The analytical predictions from Eq.~\eqref{eq:sensitivity_factor} (solid curves) are superimposed on the numerical data and show excellent agreement, consistent with Fig.~\ref{fig_app:freq_amplitude_theo_numerics} in Appendix~\ref{app_sec:muliple_scale}

\section{\label{sec:stability_breathers}Linear stability and dynamics}

\begin{figure*}[!t]
    \centering
    \includegraphics[width=0.8\textwidth]{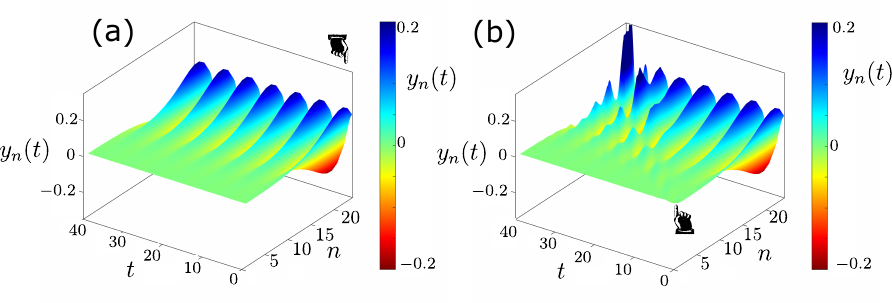}
    \caption{
        Spatiotemporal evolution of the displacement of a perturbed, linearly stable nonlinear skin breathing mode.
        The unperturbed mode has amplitude $\lVert \vec{y}\rVert^{2} = 0.2$ and belongs to the family originating from the skin mode with $j = 1$ in the case of softening nonlinearity, $g=-1$ [see Figs.~\ref{fig:nonlinear_continuation_results}(e) and~\ref{fig:floquet_analysis}(c)].
        Here $N = 24$, $\gamma = 0.25$, and $\alpha = 1$.
        The initial conditions for the time evolution are $y_n(t=0) + \delta z_n(t=0)$ with $\delta = 0.005$ and $\lVert \vec{z}\rVert = 1$, see text for details.
        In (a), the perturbation is applied at the end favored by the NHSE, $\vec{z} = (0, 0, \ldots, 0, 1)^{T}$, with all momenta $\dot{z}_n = 0$ while in (b) a similar perturbation is applied at the opposite end with $\vec{z} = (1, 0, \ldots, 0, 0)^{T}$, see hand symbols.
        In (a) the perturbed linearly stable solution exhibits stable dynamics, whereas (b) shows the onset of instability.
    }
    \label{fig:dynamics_perturbed_breather}
\end{figure*}

We now turn to the investigation of the linear stability of the nonlinear skin breathing modes presented in Sec.~\ref{sec:skin_breathers}.
Substituting $y_n + \delta z_n$ ($\delta \ll 1$) within the equations of motion [Eq.~\eqref{eq:eq_motion}]  and neglecting the higher order terms in $\delta$, leads to the variational equations
\begin{equation}
    \ddot{z}_n = a_{-}z_{n-1} + a_{+}z_{n+1} - (2 + \alpha) z_n - 3g\left[y_n^2(t)\right]z_n,
    \label{eq:eq_perturbation}
\end{equation}
with $\vec{z}=\left(z_1, z_2, \ldots, z_N, \dot{z}_1, \ldots, \dot{z}_N\right)^{T}$.
In the general setting of first-order differential equations, the variational equations reduce to
\begin{equation}
    \dot{\vec{x}} = Z[\vec{y}(t)] \vec{x},
    \label{eq:first_order_variational}
\end{equation}
which generates the time evolution of the deviation vector $\vec{x}=\left(x_1, x_2, \ldots, x_N, x_{N+1}, \ldots, x_{2N}\right)^{T} = \left(z_1, z_2, \ldots, z_N, \dot{z}_1, \ldots, \dot{z}_N\right)^{T}$.
Note that in Eq.~\eqref{eq:first_order_variational}, $Z$ is a linear, non-Hermitian matrix with time-periodic coefficients; see Appendix~\ref{app_sec:matrices}.
It follows that we need to rely on Floquet theory~\cite{FG2008} to study the linear stability of $y_n(t)$ through the properties of $Z$.
Using the similarity transformation, $x_n = r^{n/2}\eta_n$, we map the variational equations above to a Hermitian representation~\cite{KSUS2019},
\begin{equation}
    \dot{\vec{\eta}} = W [y_n(t)] \vec{\eta},
    \label{eq:first_order_variational_hermitian}
\end{equation}
with $\vec{\eta}=\left(\eta_1, \eta_2, \ldots, \eta_N, \eta_{N+1}, \ldots, \eta_{2N}\right)^{T}$, and the $W$ is shown in Appendix~\ref{app_sec:matrices}.
Here the $W$ is a linear, Hermitian and symplectic matrix (see Appendix~\ref{app_sec:matrices}) with real coefficients and shares the same eigenvalues as $Z$~\cite{KEBB2018}.

More specifically, the integration over a period $T_b=2\pi/\Omega$ of the variational equations starting from $2\times N$ initial deviation vectors $\vec{\eta}(t=0)$ forming a basis in the tangent space leads to
\begin{equation}
    \vec{\eta} (T_b) = \Phi(T_b) \vec{\eta}(0),
    \label{eq:monodromy}
\end{equation}
where $\Phi$ is the so-called monodromy matrix.
The $\Phi$ shares the same spectral properties as the $W [y_n(t)]$. 
It follows that if $\lambda$ is a Floquet eigenvalue of $\Phi$, $\lambda^{\star}$, $1/\lambda$ and $1/\lambda^{\star}$ are also Floquet eigenvalues~\cite{A1997,MA1998,S2001,FG2008}.
Accordingly, the breathing solution is linearly stable if $\lvert \lambda \rvert = 1$, and unstable otherwise [see Eq.~\eqref{eq:monodromy}].

It is also instructive to examine the shape of the Floquet eigenvectors.
Equation~\eqref{eq:monodromy} gives, $\eta_n(t) = \zeta_n(t) \lambda^{t/T_b}$, where $\zeta_n(t)$ is the eigenvector associated with $\lambda$.
Because of the skin profile of $y_n(t)$, the sites near the left boundary remain effectively at rest, $y_n(t)\to 0$ ($T_b\rightarrow \infty$).
In this region, the corresponding time evolution of the perturbation $\eta_n(t) = \zeta_n(t)$ can be found by noticing that Eq.~\eqref{eq:first_order_variational_hermitian} reduces to an eigenvalue problem, when looking for solutions of the form, $\zeta_n (t) = \zeta_n e^{i\theta t} $.
Thus the eigenvector $\eta_n = \zeta_n \propto \sin \left( nj\pi/N\right)$ are spatially extended and symmetric, leading to the deviation vector $x_n(t) = r^{n/2} \zeta_n(t) \lambda^{t/T_b}$ within the non-Hermitian framework of $y_n$ leaning in the direction favored by the NHSE.

Figure~\ref{fig:floquet_analysis}(a) shows the numerically computed dependence of the magnitude of the Floquet eigenvalues, $|\lambda|$, on the amplitude $\lVert \vec{y}\rVert^{2}$ for the family emerging from the skin mode with $j=1$, assuming $N=24$, $\alpha = 1$, $\gamma = 0.25$, and $g = -1$ [see the red curve in Fig.~\ref{fig:nonlinear_continuation_results}(d)].
We find that nonlinear skin breathing modes are linearly stable at weak amplitudes, supported by all Floquet eigenvalues satisfying $\lvert\lambda\rvert = 1$. An example of such a stable mode, together with its associated dynamics, are shown in Figs.~\ref{fig:nonlinear_continuation_results}(e)–(f) [see also Figs.~\ref{fig:nonlinear_continuation_results}(b)–(c) for the case $g=1$].
Upon further increasing the amplitude, the nonlinear skin breathing modes typically become unstable, as evidenced by the appearance of Floquet eigenvalues whose moduli deviate from unity, i.e., $|\lambda|\neq 1$.
Following such scenario, Figs.~\ref{fig:floquet_analysis}(c) and (d) display in the complex plane representative sets of Floquet eigenvalues at $\lVert \vec{y}\rVert^{2} = 1$ (purple dashed horizontal line) and $\lVert \vec{y}\rVert^{2} = 3$ (green dotted horizontal line).
We see that at $\lVert \vec{y}\rVert^{2} = 1$, the nonlinear skin mode manifests a real instability [Fig.~\ref{fig:floquet_analysis}(c)], whereas the larger-amplitude mode with $\lVert \vec{y}\rVert^{2} = 3$ displays both real and complex instabilities [Fig.~\ref{fig:floquet_analysis}(d)].

In Figs.~\ref{fig:nonlinear_continuation_results}(a) and (d), we superimpose the results of the linear stability analysis on top of the $\Omega$ versus $\lVert \vec{y}\rVert^{2}$ curves for the presented families of nonlinear skin breathing modes.
Stable branches are shown as solid curves, whereas unstable segments are indicated by dashed ones.
Overall, we find that nonlinear skin breathing modes at weak amplitudes are generally stable, while those at moderate amplitudes may show alternating stable and unstable regions, consistent with the behavior illustrated in Fig.~\ref{fig:floquet_analysis}(a).

Figure~\ref{fig:floquet_analysis}(d) depicts the Floquet eigenvectors associated with the eigenvalues presented in Fig.~\ref{fig:floquet_analysis}(c).
These eigenvectors are ordered by increasing magnitude of their corresponding Floquet eigenvalues.
They clearly exhibit skinny-like spatial profiles, reflecting the characteristic boundary localization induced by nonreciprocity.
Importantly, these Floquet eigenvectors provide the generic form of perturbations, $\delta z_n$, that determine the linear stability of the nonlinear skin breathing modes.
It follows that this stability feature appears to be a robust and persistent property of both stationary and time-periodic nonlinear normal modes in nonreciprocal chains; see also Ref.~\cite{MCKA2024}.

To prove the above argument, we perform direct numerical simulations of the equations of motion [Eq.~\eqref{eq:eq_motion}], using as initial condition a linearly stable nonlinear skin breathing mode with amplitude $\lVert \vec{y} \rVert^{2} = 0.2$, shown in Fig.~\ref{fig:nonlinear_continuation_results}(e).
We perturb this solution with a deviation vector, $\vec{z}(0) = (z_1=0, \ldots, z_N = 1, \dot{z}_1 = 0, \ldots, \dot{z}_N=0)^T$, localized at the (right) end favored by the NHSE, with magnitude $\delta \approx 0.005$ [see hand mark in Fig.~\ref{fig:dynamics_perturbed_breather}(a)].
The result of this simulation in Fig.~\ref{fig:dynamics_perturbed_breather}(a) clearly displays a coherent and oscillating envelope that persists up to the final simulation time $t = 40$.
In contrast, the dynamics of the same linearly stable breathing mode, when perturbed at the (left) end opposite to the NHSE-favored side with $\vec{z}(0) = (z_1=1, \ldots, z_N = 0, \dot{z}_1 = 0, \ldots, \dot{z}_N=0)^T$ [see hand mark in Fig.~\ref{fig:dynamics_perturbed_breather}(b)], exhibits incoherent behavior with wave amplification, as if the mode were linearly unstable.

To understand this outcome, we project the perturbation $\vec{z}(t = 0) = (1, 0, \ldots, 0)^{T}$ onto the normal modes of the system, $\delta z_n = \sum_j c_j u_{j,n}$, assuming a bi-orthonormalization; see Eq.~\eqref{eq:normal_modes}.
We find that $ c_j = 2 \delta r^{-1/2} \sin \left[j\pi/(N+1)\right]$, implying that the perturbation strongly excites all the normal modes, since $\lvert c_j\rvert /\max \lvert c_j\rvert \sim 1$.
It follows that, $\vec{z}(0)$ grows rapidly in time due to the NHSE and interacts with the stable nonlinear breathing mode at the right edge of the chain, leading to its destabilization and the unstable dynamics observed in Fig.~\ref{fig:dynamics_perturbed_breather}(b).

\section{\label{sec:conclusion_breathers}Conclusion and outlooks}
We have confirmed the existence of a nonlinear non-Hermitian skin effect (NHSE) in nonreciprocal mechanical lattices.
Using a combination of multiple-scale analysis and numerical pseudo-arclength continuation methods, we identified families of nonlinear skin breathing modes that emerge from the linear skin modes at finite amplitudes, for both hardening and softening nonlinearities.
These nonlinear skin breathing modes remain spatially localized at the edge selected by the NHSE and exhibit time-oscillating envelopes, with a nonlinear frequency that depends on the amplitude, the nonreciprocity, and the lattice size.

These results extend the nonlinear NHSE from stationary waves to time-periodic ones, thereby opening new opportunities for wave manipulation in nonlinear mechanical metamaterials. 
Looking ahead, several questions naturally arise.
The first direction concerns the throughout investigation of bulk breathers in the setting above, and the conditions under which they may become mobile~\cite{VGBVTCC2025}.
Another promising avenue is to explore the interplay between the nonlinear skin breathing states discussed above and the edge modes in topological mechanical lattices~\cite{SBPM2024}.
Finally, extending these results to higher-dimensional systems would also be an interesting direction for future work.

\begin{acknowledgments}
B.M.M acknowledges partial support from the Israel Science Foundation (ISF) and from the EU H2020 ERC StG ``NASA'' Grant Agreement No.~101077954.
B.M.M is grateful to Vassos Achilleos and Georgios Theocharis for insightful discussions. 
The authors also thank the anonymous reviewers for their constructive comments and suggestions, which significantly improved the clarity of the manuscript.
\end{acknowledgments}

\appendix

\section{\label{app_sec:matrices}Some useful matrices}
In this section, we give the explicit formula for some of the matrices used in the main text.
\subsection{\label{app_subsec:dynamical_matrix}Dynamical matrix and its properties}
First is the dynamical operator, $D$ [Eq.~\eqref{eq:dynamical_matrix}] which is a $N\times N$ matrix reading 
\begin{equation}
    D = \begin{pmatrix}
           2 + \alpha & - a_{+} & 0 & \ldots & 0 & 0\\
           -a_{-} & 2 + \alpha &- a_{+} & \ldots & 0 & 0 \\
           0 & -a_{-} & 2 + \alpha & \ldots & 0 & 0 \\
           \vdots & \vdots & \vdots & \ddots & \vdots & \vdots \\
           0 & 0 & 0 & \ldots & 2 + \alpha & - a_{+} \\
           0 & 0 & 0 & \ldots & -a_{-} & 2 + \alpha \\
       \end{pmatrix}.
       \label{eq:entries_of_dynamical_matrix}
\end{equation}
The $a_{\pm} = 1 \mp \gamma$ ensure that the dynamical matrix has constant diagonal entries.
Clearly, the coefficients of the dynamical matrix are real.
Further, the Hermitian conjugate matrix
\begin{equation} 
    D^\dagger = 
    \begin{pmatrix} 2 + \alpha & - a_{-} & 0 & \ldots & 0 & 0\\ 
    -a_{+} & 2 + \alpha &- a_{-} & \ldots & 0 & 0 \\ 
    0 & -a_{+} & 2 + \alpha & \ldots & 0 & 0 \\ 
    \vdots & \vdots & \vdots & \ddots & \vdots & \vdots \\ 
    0 & 0 & 0 & \ldots & 2 + \alpha & - a_{-} \\ 
    0 & 0 & 0 & \ldots & -a_{+} & 2 + \alpha \\ 
    \end{pmatrix}, 
    \label{eq:entries_of_dynamical_matrix} 
\end{equation}
with ${^\dagger}$ denoting the conjugate transpose operation~\cite{HN1996,BB1998,AGU2020}.
Consequently $D^\dagger \neq D$ and therefore $D$ is non-Hermitian, as stated in the main text.

\subsection{\label{app_subsec:stability_matrix} Stability matrix of the variational equations}

The next operator $Z$ [Eq.~\eqref{eq:first_order_variational}] is a $2N\times 2N$ matrix obtained when expressing the variational equations, as a first order differential equations with the non-Hermitian variables, $x_n$.
Its expression reads
\begin{equation}
    Z = 
    \begin{pmatrix}
        0_N & I_N \\ 
        Z_{2,1} & 0_N
    \end{pmatrix},
\end{equation}
with $0_N$ and $I_N$ being the null and identity matrices of dimensions $N$ and 
\begin{widetext}
    \begin{equation}
    Z_{2,1} = \begin{pmatrix}
           -2 - \alpha - 3g y_1^2 &  a_{+} & 0 & \ldots & 0 & 0\\
          a_{-} & -2 - \alpha - 3g y_2^2 & a_{+} & \ldots & 0 & 0 \\
           0 & a_{-} & -2 - \alpha - 3g y_3^2 & \ldots & 0 & 0 \\
           \vdots & \vdots & \vdots & \ddots & \vdots & \vdots \\
           0 & 0 & 0 & \ldots & -2 - \alpha - 3g y_{N-1}^2 & a_{+} \\
           0 & 0 & 0 & \ldots & a_{-} & -2 - \alpha - 3g y_N^2 \\
       \end{pmatrix}.
        \label{eq:z21}
\end{equation}
\end{widetext}
This matrix can be rewritten as 
\begin{equation}
    Z = JU, \qquad U^T \neq U,
\end{equation}
where
\begin{equation}
    J = 
    \begin{pmatrix}
        0_N & -I_N \\ 
        I_N & 0_N
    \end{pmatrix},
\end{equation}
is the symplectic matrix.
Then, we can easily check the symplectic condition,
\begin{equation}
    Z^TJZ=J, \quad \mbox{or equiv.} \quad U^{T}JU = J,
\end{equation}
which holds whenever $U^{T}=U$.
In the non-Hermitian case, $U$ is not symplectic.

Equivalently in the Hermitian framework with variables $\eta_n$ [see Eq.~\eqref{eq:first_order_variational_hermitian}] we express the $2N\times 2N$ matrix 
\begin{equation}
    W = 
    \begin{pmatrix}
        0_N & I_N \\ 
        W_{2,1} & 0_N
    \end{pmatrix}.
\end{equation}
The expression of $W_{2,1}$ is similar to $Z_{2,1}$~\eqref{eq:z21}, except taht the upper and lower diagonals have equal entries $[W_{2,1}]_{n,n+1}=\sqrt{a_{+}a_{-}}$ and $[W_{2,1}]_{n+1,n}=\sqrt{a_{+}a_{-}}$.
On the other hand, the diagonal entries reads $[W_{2,1}]_{n,n}=-2-\alpha - 3gy_n^2$.
As such, rewriting $W=JV$, it is easy to see that $V^{T} = V$ since $V$ is Hermitian, symplectic with real entries.
It follows that the Floquet eigenvalues can be straightforwardly characterized in this setting, as discussed in the main text, see also~\cite{A1997,FW1998,MA1998}.

\section{\label{app_sec:muliple_scale}Multiple scale analysis}

The equations of motion of the nonreciprocal Klein-Gordon chain of nonlinear classical oscillators read
\begin{equation}
\ddot{y}_n = a{-}\left(y_{n-1} - y_n\right) + a_{+}\left(y_{n+1} - y_n\right) - \alpha y_n - g y_n^3,
\label{eq_app:eq_motion}
\end{equation}
where $\vec{y} = (y_1, y_2, \ldots, y_n, \ldots, y_{N-1}, y_N)^T$ is the displacement vector, $N$ is the chain size, $a_{\pm} = 1 \mp \gamma$, $\alpha > 0$, and $g = \pm 1$. Throughout, we consider fixed boundary conditions, i.e. $y_0 = y_{N+1}=0$.

This equation can be reduced to
\begin{equation}
\ddot{y}_n =  -\sum_m D_{n,m} y_m - g y_n^3,
\label{eq_app:eq_motion_reduced}
\end{equation}
where the matrix $D$ is given in Sec.~\ref{app_subsec:dynamical_matrix}.
In the linear limit, $y_n \rightarrow 0$, we look for solutions of the form $y_n(t) = u_n e^{i\omega t} + \text{c.c.}$, which lead to the eigenvalue problem $\omega^2 \vec{u} = D \vec{u}$.
We obtain $N$ frequencies $\omega_j$ ($j = 1, 2, \ldots, N$), associated with $N$ right (left) eigenvectors $\vec{u}_j$ ($\vec{v}_j$).
Their expressions are given in Eqs.~\eqref{eq:linear_square_frequency} and \eqref{eq:normal_modes} of the main text.

Our goal is to show that nonlinear normal modes emerge from $\vec{u}_j$, satisfying Eq.~\eqref{eq_app:eq_motion_reduced}.
We apply the multiple-scale expansion, assuming a solution of the form~\cite{NM1979}
\begin{equation}
y_n(t, T) = \epsilon \left( y_n^{(1)} + \epsilon^2 y_n^{(3)} + \ldots \right),
\label{eq_app:ansatz_perturb_01}
\end{equation}
where $t$ and $T = \epsilon^2 t$ are the fast and slow time variables, respectively.
It follows that
\begin{equation}
\frac{d^2}{dt^2}
= \left( \frac{\partial}{\partial t} + \epsilon^2 \frac{\partial}{\partial T} \right)^2.
\label{eq_app:derivs_time_variables}
\end{equation}
In what follows, we denote $\partial/\partial t \rightarrow \partial_t$ and $\partial/\partial T \rightarrow \partial_T$.

Substituting Eq.~\eqref{eq_app:derivs_time_variables} into Eq.~\eqref{eq_app:eq_motion_reduced}, we obtain
\begin{equation}
    \begin{split}
        \partial_{tt}^2\left(\epsilon y_n^{(1)} +  \epsilon^3 y_n^{(3)}\right)  + 2\epsilon \partial^2_{Tt}\left(\epsilon y_n^{(1)} + \epsilon^3 y_n^{(3)}\right) = \\
        -\epsilon \sum_n D_{n, m} y_n^{(1)} - \epsilon^3 \sum_n D_{n, m} y_n^{(3)} - g \epsilon^3 \left[ y_n^{(1)}  \right]^3,
    \end{split}
\end{equation}
where we have neglected terms of order $\epsilon^4$, $\epsilon^5$ and higher.

Collecting the terms at order $\epsilon$, we obtain
\begin{equation}
\partial_{tt}^2 y_n^{(1)} = - \sum_m D_{n,m} y_m^{(1)}.
\label{eq_app:first_order_perturb_01}
\end{equation}
Thus, we seek solutions of the form
\begin{equation}
    y_n^{(1)} = A(T) w_n e^{i\Omega t} + A^\star(T) w_n e^{-i\Omega t},
    \label{eq_app:ansatz_first_order_perturb}
\end{equation}
with $\{^\star\}$ being the complex conjugate operation.
This leads to the eigenvalue problem
\begin{equation}
    \Omega^2 w_n = \sum_m D_{n,m}w_m,
    \label{eq_app:first_order_perturb_02}
\end{equation}
whose solutions are the frequencies 
\begin{align}
    \Omega_j^2 = \omega_j^2 &= 2 + \alpha - 2\sqrt{1 - \gamma^{2}} \cos\left(\frac{j\pi}{N+1}\right), 
\end{align}
and eigenvectors (skin modes),
\begin{align}
    w_{j,n} = u_{j,n} &= r^{-n/2} \sin\left(\frac{n j \pi}{N+1}\right).
\end{align}
 of the linearized limit.
For simplicity, in what follows we omit the index $j$.

\begin{figure}[!tb]
    \centering
    \includegraphics[width=\columnwidth]{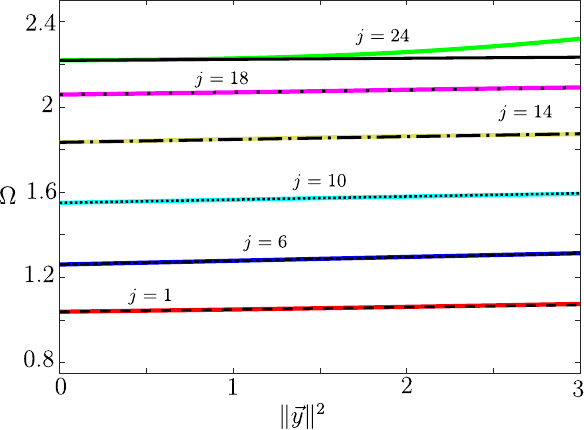}
    \caption{
        Comparison of theoretical and numerical results for the case $g=1$.
        Dependence of the numerically obtained nonlinear frequency $\Omega$ against the amplitude $\lVert \vec{y} \rVert^{2}$ for the nonlinear breathing modes with $j=1$ (red), $j=6$ (blue), $j=10$ (cyan), $j=14$ (yellow), $j=18$ (purple), and $j=24$ (green) from the bottom to the top curves respectively.
        The analytical predictions obtained from the multiple-scale analysis for these representative modes are superimposed as black curves for the case with $j=1$ (dashed-dashed), $j=6$ (dashed-space), $j=10$ (dotted-dotted), $j=14$ (dashed-dotted), $j=18$ (dotted-space) and $j=24$ (solid).
    }
    \label{fig_app:freq_amplitude_theo_numerics}
\end{figure}

Proceeding to order $\epsilon^3$, we find that
\begin{equation}
\partial_{tt}^2 y_n^{(3)} + \sum_{m} D_{n,m} y_m^{(3)}
= - 2 \partial_{Tt}^2 y_n^{(1)} - g \left[ y_n^{(1)} \right]^3,
\label{eq_app:second_order_perturb}
\end{equation}
with
\begin{multline*}
    \left[ y_n^{(1)} \right]^3
    = 3 \lvert A\rvert^2 A (u_n)^2 u_n e^{i\omega t}
    + 3 \lvert A \rvert^2 A^\star (u_n)^2 u_n e^{-i\omega t} 
    \\[2mm]
    + A^3 (u_n)^3 e^{3 i \omega t}
    + A^{\star 3} (u_n)^3 e^{-3 i \omega t}.
\end{multline*}
It follows that when canceling the resonant terms, we obtain
\begin{equation}
    - 2 i \omega u_n e^{i\omega t}\partial_T A
    - 3 g  \lvert A\rvert^2 A (u_n)^2 u_ne^{i\omega t} = 0.
\end{equation}

Assuming a bi-orthogonal set of right and left eigenvectors
$\vec{u}$ and $\vec{v}$ satisfying $\vec{v}_j^\star \vec{u}_l = 0$ if $j\neq l$~\cite{AGU2020},
we multiply the above expression from the left by
$\vec{v}^{\star} e^{-i\omega t}$.
Explicitly, this operation reads in its matrix form,
\begin{widetext}
    \begin{equation}
   \begin{pmatrix}
         v_1e^{-i\omega t}, & v_2e^{-i\omega t}, &   v_3e^{-i\omega t} & \ldots, v_Ne^{-i\omega t}
    \end{pmatrix} 
    \times 
    \left[
        -2i \omega
        \begin{pmatrix}
            u_1 e^{i\omega t} \\ u_2e^{i\omega t} \\ u_3e^{i\omega t} \\ \vdots \\ u_Ne^{i\omega t}
        \end{pmatrix}
        \partial_T A - 3 g \lvert A\rvert^2 A
        \begin{pmatrix}
            \left(u_1\right)^2 u_1e^{i\omega t} \\ \left(u_2\right)^2 u_2e^{i\omega t} \\ \left(u_3\right)^2 u_3e^{i\omega t} \\ \vdots \\ \left(u_N\right)^2 u_Ne^{i\omega t}
        \end{pmatrix}
    \right] = 0.
\end{equation}
\end{widetext}

Consequently,
\begin{equation}
        2 i \omega \partial_T A
        + 3 \mu \lvert A\rvert^2 A = 0,
        \label{eq_app:dnls}
\end{equation}
with $\mu = g \displaystyle\sum_{n=1}^N v_n u_n^3/\displaystyle\sum_{n=1}^N v_n u_n$.

Thus looking for a slowly varying envelope in the form
\begin{equation}
    A(T) = A_0 e^{-i\nu T},
\end{equation}
leads to
\begin{equation}
    \nu = - \frac{3\mu \lvert A_0\rvert^2}{2\omega}.
\end{equation}
Consequently,
\begin{equation}
    A(T) = A_0 e^{iT \frac{3\mu \lvert A_0 \rvert^2}{2\omega}}.
    \label{eq_app:amplitude_dnls}
\end{equation}
By substituting Eq.~\eqref{eq_app:amplitude_dnls} into Eq.~\eqref{eq_app:ansatz_first_order_perturb}, and then into Eq.~\eqref{eq_app:ansatz_perturb_01}, and considering terms at order $\epsilon$, we obtain
\begin{equation}
    y_n(t) = 2 \epsilon A_0 u_n \cos\left( \Omega t \right),
    \label{eq_app:nonlinear_modes_01}
\end{equation}
where
\begin{equation}
    \Omega = \omega + \frac{3 \epsilon^{2} \mu \lvert A_0 \rvert^2}{2\omega}.
    \label{eq_app:nonlinear_freq_01}
\end{equation}

From Eq.~\eqref{eq_app:ansatz_perturb_01}, when the phase of the solution is such that $\dot{y}_n = 0$, the maximal displacement at site $n$ is given by $y_n = 2\,\epsilon A_0 u_n$.
It follows that,
\begin{equation}
    \lVert \vec{y} \rVert^{2}
    = 4 \epsilon^{2} \lvert A_0 \rvert^2\sum_{n} u_n^{2},
    \label{eq_app:expressing_epsilon}
\end{equation}
is a constant coefficient, also due to the fact that $\lvert A\rvert^2 = \lvert A_0\rvert^2$ is a conserved quantity of the the amplitude equation, see also Appendix~\ref{app_sec:conservation}.
Substituting Eq.~\eqref{eq_app:expressing_epsilon} into Eqs.~\eqref{eq_app:nonlinear_modes_01} and \eqref{eq_app:nonlinear_freq_01}, it follows that
\begin{equation}
    y_n(t) = \sqrt{S}\left[ \frac{u_n}{\lVert \vec{u} \rVert} \right]
    \cos\left( \Omega t \right),
    \qquad S = \lVert \vec{y} \rVert^{2},
    \label{eq_app:nonlinear_modes_02}
\end{equation}
and
\begin{equation}
    \Omega = \omega
    + \frac{3 S g}{8 \omega}
    \left[\frac{\displaystyle \sum_{n=1}^{N} v_n u_n^{3}}
    {\displaystyle
    \left(\sum_{n=1}^{N} u_n^{2}\right)
    \left(\sum_{n=1}^{N} v_n u_n\right)}
    \right].
    \label{eq_app:nonlinear_freq_02}
\end{equation}
These are the expressions used in the main text, with $\sum_{n=1}^{N} v_n u_n = (N+1)/2$.

Figure~\ref{fig_app:freq_amplitude_theo_numerics} shows the dependence of the nonlinear frequency $\Omega$ on the amplitude, $\lVert \vec{y}\rVert^2$, obtained via numerical continuation for a lattice with $N=24$, $\alpha = 1$, $\gamma =0.25$ and $g = 1$ [see also Fig.~\ref{fig:nonlinear_continuation_results}(b)].
This dependence is computed for representative families of nonlinear breathing modes with $j=1$ (red), $j=6$ (blue), $j=14$ (cyan), $j=18$ (yellow), and $j=24$ (green), corresponding to the curves from bottom to top.
In addition, we superimpose the analytical predictions obtained from Eq.~\eqref{eq_app:nonlinear_freq_02}, as black curves in Fig.~\ref{fig_app:freq_amplitude_theo_numerics} for the representative families above with $j=1$ (dashed–dashed), $j=6$ (dashed–space), $j=10$ (dotted–dotted), $j=14$ (dashed–dotted), $j=18$ (dotted–space), and $j=24$ (solid).
For each of the displayed family, we observe an excellent agreement between the analytical and numerical results up to amplitudes $\lVert \vec{y} \rVert^{2} \approx 2$ and $3$.

\section{\label{app_sec:conservation}Amplitude equation and conservation law}
While nonlinear nonreciprocal Klein–Gordon chain [Eq.~\eqref{eq:eq_motion}] exhibits a non-Hermitian dynamics, the multiple-scale analysis above reveals that the complex amplitude $A(T)$, with $T=\epsilon^2 t$, satisfies at leading order the nonlinear Schr\"odinger equation,
\begin{equation}
    i\partial_T A = \kappa |A|^2 A,  \quad \kappa = -\frac{3\mu}{2\omega}.
    \label{eq:amp_eq2}
\end{equation}
whose dynamics is Hermitian.
Therefore Eq.~\eqref{eq:amp_eq2} possesses conserved quantities.

{\it Conservation of $\lvert A\rvert^2$.}
Taking the derivative of the squared modulus of the amplitude yields
\begin{equation}
    \frac{d}{dT} \lvert A\rvert^2 = (\partial_T A)A^\star + A(\partial_T  A^\star) .
    \label{eq:norm_derivative}
\end{equation}
From Eq.~\eqref{eq:amp_eq2}, we have $\partial_T A = - i \kappa \lvert A\rvert^2 A$ and $\partial_T A^\star = + i \kappa \lvert A\rvert^2 A^\star$.
Substituting these expressions into Eq.~\eqref{eq:norm_derivative} gives
\begin{equation}
    \frac{d}{dT} \lvert A\rvert^2 = (- i \kappa \lvert A\rvert^2 A) A^\star + A ( i \kappa \lvert A\rvert^2 A^\star ) = 0.
\end{equation}
Therefore,
\begin{equation}
    \frac{d}{dT} \lvert A\rvert^2 = 0 ,
\end{equation}
and the quantity $\lvert A(T)\rvert^2$ is a constant of motion.
Thus $\lvert A (T)\rvert^2 = \lvert A_0 \rvert^2$ from Eq.~\eqref{eq:breathers_expression_in_hermitian} and squared maximum displacement (also referred to as the total amplitude)
\begin{equation}
    \lVert \vec{y} \rVert^{2}
    = 4 \epsilon^{2} \lvert A_0 \rvert^2\sum_{n} u_n^{2},
\end{equation}
is a constant coefficient.


\bibliography{apssamp}

\begin{thebibliography}{54}
\bibitem [{QQQ()}]{QQQQQQQ}%
  \BibitemOpen
  \href@noop {} {}\bibinfo {note} {The associated $\sigma$ values are $\sigma \approx 0.03$, $0.02$ and $0.016$ in case of $N=24$, $32$ and $42$ respectively.}\BibitemShut {Stop}%
\end{thebibliography}

\end{document}